\documentclass[%
 reprint,
 amsmath,amssymb,
 aps,
nofootinbib]{revtex4-2}

\usepackage{graphicx}
\usepackage{dcolumn}
\usepackage{bm}
\usepackage{physics}
\usepackage{amsmath,amssymb}
\usepackage{graphicx}
\usepackage{tikz-feynman}
\usepackage{subcaption}
\usepackage[font=small]{caption}
\DeclareMathOperator*{\SumInt}{%
\mathchoice%
 {\ooalign{$\displaystyle\sum$\cr\hidewidth$\displaystyle\int$\hidewidth\cr}}
 {\ooalign{\raisebox{.14\height}{\scalebox{.7}{$\textstyle\sum$}}\cr\hidewidth$\textstyle\int$\hidewidth\cr}}
 {\ooalign{\raisebox{.2\height}{\scalebox{.6}{$\scriptstyle\sum$}}\cr$\scriptstyle\int$\cr}}
 {\ooalign{\raisebox{.2\height}{\scalebox{.6}{$\scriptstyle\sum$}}\cr$\scriptstyle\int$\cr}}
}

\begin{document}

\preprint{APS/123-QED}

\title{The Classical Gravitational Impulse at High Energies}

\author{Michael Saavedra}
\email{msaavedra@physics.ucla.edu} 
\affiliation{Mani L. Bhaumik Institute for Theoretical Physics,
 University of California at Los Angeles, Los Angeles, CA 90095, USA}
\date{\today}

\begin{abstract}
We compute the gravitational impulse for two classical massive scalars in the ultrarelativistic limit to all orders in Newton's constant $G_N$ at fixed $G_N s/m b$ to $O(m^4/s^2)$. By computing the 4 and 5-point scattering amplitudes in the small-mass regime of $-t\sim m^2$, we are able to resum all large $G_N s$ corrections. Applying the KMOC formula for the impulse and taking the large mass limit, we recover the classical result at high energies. This resummation is in complete agreement with known results in the post-Minkowski expansion, and in the massless limit we recover previous results for the radiated energy. 
We use this resummed amplitude to predict the leading high-energy behavior of the PM expansion to eleventh post-Minkowski order. 
\end{abstract}

\maketitle

\paragraph*{Introduction}

Since the discovery of gravitational waves \cite{LIGOScientific:2018mvr, LIGOScientific:2020ibl,KAGRA:2021vkt}, there has been an intense theoretical interest in the gravitational two-body problem. A number of tools have been developed to study the various aspects of this problem, including the post-Newtonian (PN) \cite{PN1, PN2, 7c44806c-bf75-3a8d-b622-7761a1e00cd9, 10.1143/PTP.50.492, Blanchet:2013haa}, post-Minkowskian (PM) \cite{Neill:2013wsa, Bjerrum-Bohr:2018xdl,Damour:2017zjx, Cheung:2018wkq, Bern:2019nnu,Kalin:2020mvi, Mogull:2020sak}, and self-force \cite{Mino:1996nk,Quinn:1996am, Poisson:2011nh, Barack:2018yvs} expansions, which may be based on classical methods or effective field theory (EFT) \cite{Goldberger:2004jt, Porto:2016pyg}. The current state-of-the-art involves partial results at 6PN \cite{Bini:2020nsb, Bini:2020hmy,Brunello:2025gpf, Blumlein:2021txj} and 5PM \cite{Driesse:2024feo,Bern:2025wyd, Driesse:2026qiz} orders. 

Despite the remarkable progress, there are still several outstanding questions in the field. One such question concerns the fate of classical gravitational scattering in the ultra-relativistic limit. Notably, calculations at 3 and 4PM \cite{Herrmann:2021tct,Dlapa:2022lmu} show that for highly boosted projectiles, the total energy radiated by gravitational waves grows much faster than the center-of-mass energy, indicating a breakdown of perturbation theory. Such a breakdown has been known since the 70s \cite{DEath:1976bbo, Kovacs:1977uw, Kovacs:1978eu}, and computations have generally been limited to the case of vanishingly weak coupling $ G \sqrt{s}/b \ll m/\sqrt{s}$ where the PM expansion is valid, for projectiles of generic masses $m_1\sim m_2 \sim m$ and center of mass energy $\sqrt{s}$.  A notable exception is \cite{DEath:1976bbo}, in which a formula for the Bondi news was determined at leading power in $m^2/s$ to all orders in $G_Ns/(mb)$ although further analytic computation, such as the radiated energy, was determined to be intractable.

To address this limitation, several works have explored techniques for computations at high energies. In \cite{Rothstein:2024nlq}, an EFT based on Soft Collinear Effective Theory (SCET) \cite{Bauer:2000ew, Bauer:2000yr, Bauer:2001yt}\footnote{SCET gravity for hard scattering was studied in \cite{Beneke:2012xa, Okui:2017all, Beneke:2021aip, Beneke:2022ehj, Beneke:2022pue}} with Glauber operators \cite{Rothstein:2016bsq, Gao:2024qsg} was developed for gravity, and it was found that there is a tower of large logarithms in the classical amplitude, generalizing the results of Amati, Ciafaloni, and Veneziano \cite{Amati:1990xe} at 3PM. The leading log at 5PM order was subsequently computed in \cite{Alessio:2025isu}, and a formalism based on shockwaves was put forth \cite{Alessio:2026bdi} to allow for small-mass projectiles. 
For classical observables, a series of papers \cite{Gruzinov:2014moa, Ciafaloni:2015xsr, Ciafaloni:2018uwe} computed the radiated energy spectrum using classical methods and amplitudes-based methods. Remarkably, their results resum an infinite tower of $t$-channel graviton exchanges, and thus are all orders in $G_N s$. Moreover, they show that the spectrum becomes non-analytic in $G_N$ as $ G_N s$ becomes very large. In a similar vein, Refs. \cite{DiVecchia:2022nna, Alessio:2024onn} have explored the radiated energy spectrum in the zero-frequency limit (ZFL), keeping full mass dependence. This allowed them to explore the transition of the observable from the perturbative regime with $ G \sqrt{s}/b \ll m/\sqrt{s}$ to the massless regime at fixed $G\sqrt{s}/b$. However, a general framework for high-energy gravitational scattering away from the ZFL is still lacking. 

In this letter, we begin to close this gap. Using EFT methods based on \cite{Rothstein:2024nlq}, we compute the gravitational impulse for small mass projectiles by summing the dominant contributions to all orders in perturbation theory. Noticing that one recovers the high-energy limit of classical gravity in the large mass limit of this theory, we are able to extract the resummed classical impulse at fixed $G_N s/(m b)$ at leading and next-to-leading order in powers of $m^2/s\ll 1$ as a convolution in impact parameter space. This allows us to extend the validity of the calculation beyond the perturbative regime of $G_N s/(m b)\ll 1$ to the strongly coupled case of $G_N s/(m b)\sim 1$. We find that the impulse is exactly the tree-level impulse at leading power in $m^2/s$. Radiative corrections enter at $O(m^2/s)$, and are entirely first order in self-force (1SF), in agreement with the analysis of \cite{Galley:2013eba}, while conservative corrections appear at fixed order through 2PM and are 0SF. We then use the resummation to compute the leading PM results for the impulse to 11PM order. 
As a by-product, we are able to reproduce the results of \cite{Gruzinov:2014moa, Ciafaloni:2015xsr, Ciafaloni:2018uwe} for the radiated energy spectrum for massless scattering. We find that while the integrands for the radiated energy spectra in the massive and massless cases are the same as $m_i\rightarrow 0$, the convolution does not commute with the massless limit, and the radiated energies in the two cases are qualitatively different. 

\paragraph*{Conventions}

We use the mostly minus flat metric convention, $\eta = \text{diag}(1,-1,-1,-1)$, and we take $\kappa^2 = 16\pi G_N$. We use dimensional regularization for the presentation of many results, with $d= 4-2\epsilon$, and $d^\prime = d-2$. Integration measures in momentum space use the notation $[d^d k]\equiv d^d k/(2\pi)^d e^{\epsilon \gamma_E} \mu^{-2\epsilon}$. We frequently use lightcone coordinates $p^\mu = (p^+, p^-, p_\perp)$, with $p^+ = \bar{n}\cdotp p$, $p^- = n\cdotp p$, $n^2 = \bar{n}^2 = 0$ and $n\cdotp \bar{n} = 2$.  It is convenient to pick $n^\mu = (1,0,0,1)$ and $\bar{n}^\mu = (1,0,0,-1)$.

\paragraph*{Massive High-Energy Scattering}

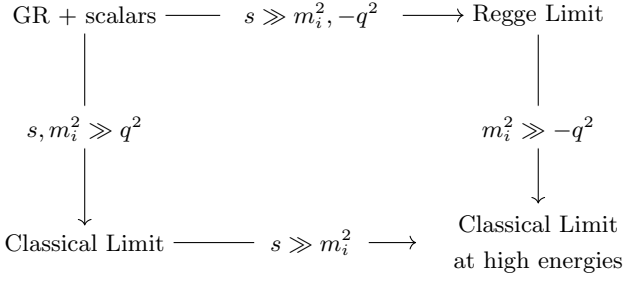
\begin{figure}
\begin{tikzpicture}
\node (a11) at (-3., 1.5) {GR + scalars};
\node (a12) at (-3., -1.5) {Classical Limit};
\node (a21) at (3., 1.5) {Regge Limit};
\node (a22) at (3., -1.25) {Classical Limit};
\node (a22) at (3., -1.75) {at high energies};
\node (a221) at (1.5, -1.5) {};
\node (a222) at (3., -1.1) {};

\draw[->] (a11) edge (a12) ;

\draw[->] (a12) edge (a221) ;

\draw[->] (a11) edge (a21);

\draw[->] (a21) edge (a222) ;

\filldraw[white] (-4,-.25) rectangle (-2,.3);
\filldraw[white] (2,-.25) rectangle (4,.3);
\filldraw[white] (-.75,-1.75) rectangle (.75,-1.2);
\filldraw[white] (-1.2,1.25) rectangle (1.2,1.8);

\node (l1) at (-3,0) {\(s, m_i^2 \gg q^2\)};
\node (l2) at (0, 1.5) {\(s\gg m_i^2, -q^2\)};
\node (l3) at (3,0) {\( m_i^2\gg -q^2\)};
\node (l4) at (0, -1.5) {\(s \gg m_i^2\)};
\end{tikzpicture}
\caption{Web of kinematic limits. The high-energy limit of classical scattering may be approached either from either the classical limit by expanding around large velocities $s/(m_1 m2)\gg 1$, or from Regge kinematics by expanding in large masses $m_i/q \gg 1$.}
\end{figure}

We consider the gravitational scattering of two massive bodies with initial momenta $p_1$ and $p_2$, with center-of-mass energy squared $s = (p_1 + p_2)^2$ and momentum transfer $q$.
In the forward limit $s\gg -q^2$, with impact parameter $b\sim q^{-1}$, one may consider different scalings of the masses with respect to the kinematics to obtain different theories. In the classical limit, one has $s\sim m^2$, with the scattering projectiles localizing onto world lines. 
The opposite limit is the small-mass regime with $m \sim q$, which is closely related to the massless case; this is the Regge limit of gravitational scattering \cite{Alessio:2026bdi}. For perturbation theory to be valid, both regimes are restricted to small-angle scattering, with angle $\theta \sim G_N \sqrt{s}/b\ll 1$.  In the massive case, there is a tighter bound, requiring $\theta \sqrt{s}/m \ll 1$ \cite{DEath:1976bbo}. The regime in which $s\gg m^2$ is particularly sensitive to this bound. This corresponds to the classical scattering of highly boosted projectiles. Traditionally,  this is studied expanding the PM result in the large boosts $\sigma\sim s/m_1 m_2$. Here, we take the alternative approach of starting with the light-mass theory and then expanding in small $q/m$ to reach this limit. In the latter case, the class of Feynman diagrams which contribute classically take a relatively simple form, allowing one to resum them to all orders in $G_N$, as discussed in Appendix \ref{Appendix_Amps}.

To describe the light-mass theory, we use a slight generalization of the EFT in \cite{Rothstein:2024nlq}, based on SCET$_{\text{M}}$\cite{Leibovich:2003jd, Rothstein:2003wh}, which will be presented in a forthcoming publication \cite{Forthcoming}. This EFT is an expansion in $\lambda^2 = -q^2/s$ and is described by collinear modes for both the gravitons and the scalars, with momenta that scale as $p_n\sim (1,\lambda^2, \lambda)$ and $p_{\bar{n}}\sim (\lambda^2,1, \lambda)$ , and soft modes with $p_S \sim (\lambda,\lambda, \lambda)$. These are then mediated by Glauber modes with $p_G \sim (\lambda^2,\lambda^2, \lambda)$, which generate the potential-like $1/p_\perp^2$ $t$-channel propagators characteristic of forward scattering. In going to the classical limit $q/m\rightarrow 0$, the scalar fields localize onto world lines, while the graviton
collinear modes become ``collinear-soft" (csoft) modes with scalings $p_{cs}\sim \lambda(\sqrt{s}/m, m/\sqrt{s},1)$ and $p_{cs}\sim \lambda( m/\sqrt{s},\sqrt{s}/m,1)$ \cite{Fleming:2007qr, Fleming:2007xt}. Taking the classical limit for the amplitudes computed in Appendix \ref{Appendix_Amps} then reduces to expanding in the region where the collinear graviton momentum has csoft scaling. 

In the massless theory, a classical contribution at $(2j + 1)$PM order scales at most as $\sim G_N s\, (G_N \sqrt{s}/b)^{2j}$, with the dominant contribution to the amplitude occurring at maximal SF order due to large logarithms \cite{Rothstein:2024nlq, Alessio:2025isu}. 
At low loop orders, the high-energy limit of the PM expansion seems to obey this scaling as well from explicit computations through 3PM. 
Starting at 4PM, however, this breaks, and one finds instead that the classical contributions instead scale as $G_N s\, (G_N \sqrt{s}/b)^j (\sqrt{s}/m)^{j-2}$, where one finds extra enhancements by the rapidities of boosted projectiles; this enhancement was first noticed at 4PM in the conservative sector in \cite{Bern:2021ppf} and later confirmed in the presence of dissipative effects in \cite{Dlapa:2022lmu}.
The origin of these enhancements can be understood light-mass theory by considering terms which appear to scale \textit{superclassically}\footnote{Formally, classical and quantum terms are not kinematically well separated in the light-mass theory, since $q\sim \hbar$ while $m\sim \hbar^0$.}, that is contributions of the form $(G_N s)^{j + 1} (G_N \sqrt{s}/b)^{2k}$. 
In passing from the light-mass theory to the classical limit, we must expand in powers of $q/m\sim 1/(m b)$, and the additional quantum suppression leads to a classical result with scaling $\sim G_N s(G_N s/(m b))^{j} (G_N \sqrt{s}/b)^{2k}$. 
Given this, we find it useful to use a power-counting in which we take the parameter $\alpha_i = G_N s/(m_i b)$ to be the coupling, and the classical result is an expansion in $\gamma^{-1} = m/\sqrt{s}\sim \sigma^{-1/2}$.
This is contrasted with both the PM expansion and the massless case, where the classical expansion parameter is taken to be $G_N \sqrt{s}/b$, and the formalism of \cite{Rothstein:2024nlq}, where the effective coupling is $G_N q^2$.

In this power-counting, each soft or collinear loop leads to $1/\gamma$ suppression in the boosted theory. 
In the light-mass theory, each soft or collinear loop scales a $G_N/b^2$. A diagram with $j$ soft/collinear loops and $k + 1$ Glauber exchanges (each of which scales as $G_N s$) then scales as $\sim (G_N/b^2)^j (G_N s)^{k+1}$, with $k\geq j$ for (super)classical scaling. Taking the classical limit, the diagram then scales as $\sim G_N s\, (m^2/s)^j \alpha^{j + k}$. Therefore, the leading contribution to a classical observable at high energies generically arises from diagrams with only Glauber exchanges, while the next-to-leading order contributions are due to diagrams with one soft or collinear loop.

\paragraph*{Gravitational Impulse}

We compute the gravitational impulse using the KMOC formalism \cite{Kosower:2018adc}. For two wave packets widely separated by an impact parameter $b$, we may compute the difference in some observable $O$ between the far past and far future as
\begin{equation}
\Delta O = \bra{\text{in}}(S^\dag O S - O)\ket{\text{in}}.
\end{equation}
For the impulse on particle $1$, the observable becomes $\mathbb{P}^\mu_1$, which picks out the momentum flowing through the massive scalar. In the classical limit, the states localize onto worldlines, and the dependence on the wave packet shape becomes a subleading effect. Then, using $S = 1 + i T$, we may write the classical impulse on particle 1 as
\begin{align}
  \Delta p_1^\mu =& \int [d^d q]\delta(q^2-2p_1\cdotp q)\delta(q^2+2p_2\cdotp q) e^{ib\cdotp q}\nonumber\\
  &\times ( I^\mu_V(q) + I^\mu_R(q)),
  \label{Wave Impulse}
\end{align}
where the momentum-space kernels are given by
\begin{align}
  I^\mu_V =& i q^\mu \mathcal{M}_{2\rightarrow 2},\nonumber\\
  I^\mu_R =& \SumInt_X \mathcal{M}^\star_{2\rightarrow 2 + X}(p_2+q,p_1-q, p_1 + \ell_1, p_2-\ell_1,\{\ell_X\}) \nonumber\\
  &\times\ell_1^\mu\, \mathcal{M}_{2\rightarrow 2 + X}(p_1,p_2, p_1 + \ell_1, p_2-\ell_1,\{\ell_X\}).
\end{align}
This is the standard KMOC formula for the impulse for the scattering of plane waves This formula reduces to that of the classical impulse in the heavy mass limit $q/m\rightarrow 0$, while maintaining manifest power-counting in the light-mass regime, so it is useful for extracting the massless and massive limits.

The outgoing projectile is on-shell and therefore must satisfy the consistency condition $(p_1 + \Delta p_1)^2 = m_1^2$.  The radiated momentum may be computed from the impulse via momentum conservation:
\begin{equation}
  R^{\mu} = -\Delta p_1^\mu - \Delta p_2^\mu.\label{Radiated Momentum}
\end{equation}
The system is symmetric under exchange of the projectiles, so $\Delta p_2^\mu$ may then be obtained from $\Delta p_1^\mu$ under the exchange $1\leftrightarrow 2$, $b\rightarrow -b$ and $n\leftrightarrow \bar{n}$.

Within the EFT framework we are working in, the sum over states implicitly contains a sum over modes, which may be soft, collinear, or anti-collinear; Glauber modes are off-shell, and thus do not appear as external states in the scattering amplitudes. 
To the order we are working at, the contributions to the impulse involve at most one soft or collinear graviton.

\paragraph*{Results}

Using the amplitudes and formula for the impulse from KMOC, we compute the eikonal and next-to-eikonal contributions to the impulse. We calculate in the full light-mass regime, and take the heavy mass limit afterwards; the four- and five-point amplitudes are given in the appendix, and the relevant diagrams are shown in Figure \ref{fig:AmpDiagrams}. To perform the computation, we Fourier transform the amplitudes to impact parameter space, so that we may trade all transverse momentum for derivatives. For the lightcone components, we use the on-shell delta functions to replace the small plus momentum $ \sim \lambda^2$ with transverse momentum and the large minus momentum $\ell^-\sim \lambda^0$.

We give all results in the center-of-mass frame and in light-cone coordinates.  Here, the incoming momenta are parameterized as $p_1^\mu = (p_1^+, \, p_1^-,\, q_\perp^\mu/2)$ and $p_2^\mu = (p_2^+, \, p_2^-,\, -q_\perp^\mu/2)$, with $q^\mu = q_\perp^\mu$ being purely transverse.  It is common in the literature to use ``dual velocity" basis $\{\check{u}_i^\mu\}$, with
\begin{equation}
    \check{u}_1^\mu = \frac{\sigma \,u_2^\mu - u_1^\mu}{\sigma^2-1},\qquad \check{u}_2^\mu = \frac{\sigma \,u_1^\mu - u_2^\mu}{\sigma^2-1},
\end{equation}
where $u_i = (p_i^+/m_i,\, p_i^-/m_i,0)$ are the velocities of the initial state projectiles, and $\sigma = u_1\cdotp u_2$.  These may be related to the light-cone coordinate basis at high energies through
\begin{align}
    \check{u}_1^\mu = \frac{m_1}{p_1^-}n^\mu - \frac{m_1 m_2^2}{s\, p_2^+}\bar{n}^\mu + O(\gamma^{-4}),
\end{align}
with the expression for $\check{u}_2^\mu$ given under the exchange $1\leftrightarrow 2$ and $n\leftrightarrow \bar{n}$.

\begin{figure}[]
  \includegraphics[width = 0.48\textwidth]{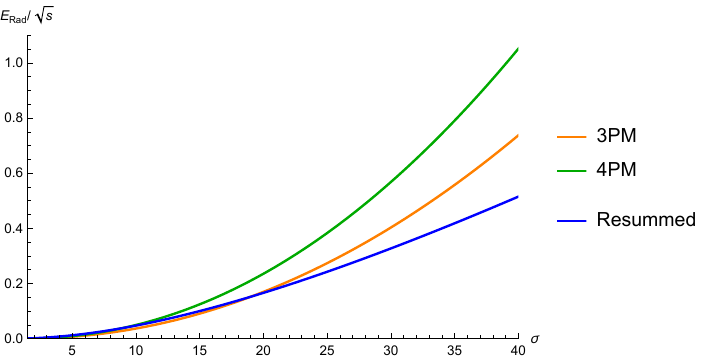}
  \caption{Total classically radiated energy at large boosts, with $G_N m /b = 1/40$, for $m_1 = m_2 = m$. The upper bound of $\sigma = 40$ corresponds to $\alpha = 2$.}
\end{figure}

In the power-counting described above, we organize the impulse in powers of $1/\gamma = m_1/p_1^-\sim \sigma^{1/2}$ as 
\begin{equation}
  \Delta p_1^\mu = \sum_{j=1}^\infty \Delta p^\mu_{1,(j)},
\end{equation}
where $\Delta p^\mu_{1,(j)}$ scales as $\sim \gamma^{-j}$.
We find the classical impulse through $O(\gamma^{-3})$ to be
\begin{align}
  \Delta p_{1,(1)}^\mu &= \partial^\mu_b\delta^{(0)},\nonumber\\
   \Delta p_{1,(2)}^\mu &= -\frac{\bar{n}^\mu}{2p_2^+}(\partial_b\delta^{(0)})^2+\frac{n^\mu}{2p_1^-}(\partial_b\delta^{(0)})^2 + \frac{\bar{n}^\mu}{2p_2^+}f^-,\label{Impulse}\\
   \Delta p_{1,(3)}^\mu &= g^\mu_\perp + f_\perp^\mu,\nonumber
\end{align}
where $\delta^{(0)}$ is the leading classical phase,
\begin{equation}
  \delta^{(0)}(b) = 4\pi\,G_N s\int\frac{[d^{d^\prime}k_\perp]}{k_\perp^2} e^{-ib\cdotp k_\perp}.
\end{equation}
The expression for $\Delta p_{1,(1)}^\mu$ and the first two terms for $\Delta p_{1,(2)}^\mu$ arise from the leading eikonal $2\rightarrow 2$ amplitude. The functions $f^+$, $g^\mu_\perp$, and $f_\perp^\mu$ arise from the next-to-eikonal amplitudes involving a collinear graviton emission. $g_\perp^\mu$ appears only at fixed order in $G_N$ and reproduces the high-energy transverse impulse at 2PM
\begin{align}
  g_\perp^\mu &= \int_{y,b^\prime} \mathcal{N}(b^\prime,m_1)\partial_b^\mu \delta^{(0)}(b-b^\prime) + (m_1\leftrightarrow m_2),\nonumber\\
  & = -\frac{15\pi}{8}\frac{G_N^2 s}{b^2}(m_1 + m_2)\, \frac{b^\mu}{|b|}+ O(\epsilon),\label{gperp}
\end{align}
where $
  \int_{y,b^\prime}\equiv \mu^{2\epsilon} \int_{-\infty}^0 dy\, \int d^{d^\prime} b$.
$\mathcal{N}(b^\prime,m_1)$ is the leading term in the series expansion in $1/\gamma$ of the numerator function $N(b^\prime)$ in Eq. (\ref{4-Point Vertex Function}), and it is given in $d=4$ as 
\begin{align}
  \mathcal{N}(b^\prime,m_1) \stackrel{d=4}{=} \frac{G_N m_1^4}{\pi^2} y^4 K_2(-y m_1 |b^\prime|)^2,
\end{align}
with $K_2$ being the modified Bessel function of the second kind. 
This, along with the other fixed-order contributions, are the only 0SF terms appearing in the impulse at this order in the $1/\gamma$ expansion.
The functions $f^+$ and $f_\perp^\mu$ meanwhile are the resummed contributions, and entirely radiative corrections. They are entirely 1SF and are given by
\begin{align}
  f^- =&- 4p_1^- p_2^+\int_{y,b^\prime}\mathcal{N}(b^\prime,m_1) \sin^2\left(\frac{y\Phi(b^\prime)}{2}\right),\nonumber\\
  f^\mu_\perp =& -2\int_{y,b^\prime}\mathcal{N}(b^\prime,m_1) \sin^2\left(\frac{y\Phi(b^\prime)}{2}\right)\nonumber\\
  & \times\partial_b^\mu \left( \delta^{(0)}(b-b^\prime) + \delta^{(0)}(b)\right),
  \label{resummed Functions}
\end{align}
with $\Phi$ being given as
\begin{align}
\Phi(b^\prime) &= -b^\prime\cdotp \partial_b\delta^{(0)}(b) -\delta^{(0)}(b-b^\prime) + \delta^{(0)}(b),\nonumber\\
& \stackrel{d=4}{=} G_N s\left( \frac{2 b^\prime\cdotp b}{b^2} + \log\frac{(b-b^\prime)^2}{b^2}\right).
\end{align} 
In general, these integrals are not analytically tractable
; however they are convergent and amenable to numerical integration.

Since $ y\sim 1/(m_1 b^\prime)$ and $\Phi \sim G_N s$, it is clear these functions depend only on the high-energy parameter $G_N s/(m b)$ (up to some overall scaling); this makes it manifestly clear that perturbation theory breaks down at ultra-relativistic energies when $G s/(m b) \sim 1$.

\paragraph*{PM Expansion to High Orders}

The resummed functions in the impulse contain information to all orders in the PM expansion at high energies. By expanding the integrals in powers of $G$, the $y$ and $b^\prime$ integrals may be performed. In particular, the $b^\prime$ integration involves only two-dimension Euclidean bubble integrals with at most a single logarithm. Using this, we predict the leading-order in $1/\gamma$ result for the plus and minus impulse at odd PM orders and the $\perp$ impulse at even PM orders through 11PM. The result for the minus impulse is
\begin{widetext}
\begin{align}
  \frac{p_2^+}{m_1^2} \Delta p_{1}^- =&-4\alpha_1^2 -\frac{35}{32}\pi\alpha_1^3(l_4+1) +\frac{63}{128}\pi\alpha_1^5\left[l_4^3 -3l_4^2 + 3 l_4 + 3\zeta_3-2\right] - \frac{1485}{2048}\pi\alpha_1^7\bigg[\frac{1}{2}l_4^5 -\frac{25}{6}l_4^4 + \frac{55}{9}l_4^3 \nonumber\\
  &+ \left(15\zeta_3+\frac{10}{3}\right)l_4^2 -\left(50\zeta_3 + \frac{25}{6}\right)l_4+\frac{45}{2}\zeta_5+\frac{55}{3}\zeta_3 + \frac{23}{6}\bigg] +\frac{5005}{16384}\pi\alpha_1^9\bigg[l_4^7 -\frac{217}{15}l_4^6 + \frac{3731}{75}l_4^5 \nonumber\\
  &+ \left(105 \zeta_3-\frac{406}{15}\right)l_4^4 - \left(868\zeta_3 - \frac{217}{15}\right)l_4^3 + \left(945 \zeta_5 + \frac{7462}{5}\zeta_3-\frac{133}{5}\right)l_4^2 + \left(640\zeta_3^2-3906\zeta_5 -\frac{1624}{5}\zeta_3 + 49\right)l_4\nonumber\\
  & + \frac{2835}{2}\zeta_7-1302\zeta_3^2 + \frac{11193}{5}\zeta_5 + \frac{217}{5}\zeta_3-48\bigg] + \frac{143325}{65536}\pi \alpha_1^{11}\bigg[-\frac18l_4^9 + \frac{741}{280}l_4^8 -\frac{38327}{2450}l_4^7 + \left(-\frac{63}{2}\zeta_3 + \frac{4561}{175} \right)l_4^6\nonumber\\
  & + \left(\frac{2223}{5}\zeta_3 - \frac{7017}{700}\right)l_4^5 + \left(-\frac{2835}{4}\zeta_5 -\frac{114981}{70}\zeta_3 -\frac{537}{140}\right)l_4^4 + \left(-945\zeta_3^2 + 6669\zeta_5 + \frac{54732}{35}\zeta_3 + \frac{129}{14}\right)l_4^3\nonumber\\
  & + \left(-\frac{25515}{4}\zeta_7 + 6669\zeta_3^2 -\frac{1034829}{70}\zeta_5 - \frac{21051}{70}\zeta_3 - \frac{888}{35}\right)l_4^2 + \bigg(-8505\zeta_5\zeta_3 + \frac{60021}{2}\zeta_7 - \frac{344943}{35}\zeta_3^2\nonumber\\
  &+ \frac{246294}{35}\zeta_5 - \frac{1611}{35}\zeta_3 + \frac{13901}{280}\bigg)l_4 -\frac{80325}{8}\zeta_9 -945\zeta_3^3 + 20007\zeta_5\zeta_3 - \frac{3104487}{140}\zeta_7 + \frac{82098}{35}\zeta_3^2 - \frac{63153}{140}\zeta_5\nonumber\\
  &+ \frac{387}{14}\zeta_3 - \frac{207181}{4200}\bigg] + O(\alpha_1^{13}).
\end{align}
\end{widetext}
In the above, we have used $l_4 = \log(4)$. We note some curious properties of the results, such as the absence of even $\zeta$ values, and that the maximal transcendental value is equal to the loop order. These match the known results for the impulse through 4PM order\footnote{Although there are results for the impulse at 5PM 1SF \cite{Driesse:2024feo}, the function basis makes it difficult to expand the results in the UR limit. We thank Jan Plefka for commenting on this matter.}.  The results for the other components of the impusle are given in Appendix \ref{B}.  

\paragraph*{Cross-Checks}

An important cross-check on this computation is that the outgoing state must be on-shell, that is the relation $(p_1 + \Delta p_1)^2 = m_1^2$ must be satisfied. Expanding in powers of $1/\gamma$, we find that this relationship is satisfied somewhat trivially through $O(\gamma^{-3})$, as the resummed functions do do appear until $O(\gamma^{-4})$. We cannot compute the full $O(\gamma^{-4})$ impulse, as this requires including subleading operators into the EFT. However, we can give a partial result for the impulse at this order, which turns out to be enough to check the on-shell conditions. Using the same strategy as above, the $O(\gamma^{-4})$ impulse may be written as
\begin{equation}
   \Delta p_{1,(4)}^\mu =\left[
   \Delta p^+_{1,SL} + 
   g^+ + f^+\right]\frac{n^\mu}{2} + \Delta p^-_{1,SL} \frac{\bar{n}^\mu}{2}.\label{N3LO Impulse}
\end{equation}
Here $\Delta p_{1,SL}^\pm$ are the terms in the impulse which are not computed using the four- and five-point amplitudes in Eq. (\ref{4-Point}) and (\ref{5-point}), and require either diagrams involving two collinear gravitons or subleading operators in the EFT. $g^+$ and $f^+$ can be computed explicitly from the amplitudes given in Appendix \ref{Appendix_Amps}, and they are given as
\begin{align}
  g^+ =& \frac{2}{p_1^-}\int_{y,b^\prime} \mathcal{N}(b^\prime,m_1)\left[\partial_{b\mu}\delta^{(0)}(b) \right]\left[\partial_{b}^\mu\delta^{(0)}(b-b^\prime)\right]\nonumber\\
  &\qquad  + (m_1 \leftrightarrow m2)\nonumber\\
  f^+ =& \frac{4}{p_1^-}\int_{y,b^\prime} \mathcal{N}(b^\prime,m_1)\sin^2\left(\frac{y \Phi(b^\prime)}{2}\right)\\
  &\qquad\times \bigg( m_1^2 +\left[\partial_{b\mu}\delta^{(0)}(b) \right] \left[\partial_{b}^\mu\delta^{(0)}(b-b^\prime)\right]\bigg).\nonumber
\end{align}
With this, we find that the on-shell condition is given as
\begin{align}
  p_1^-\Delta p^+_{1,SL} + \frac{m_1^2}{p_2^+ p_1^-}(\partial_b\delta^{(0)})^2 - \frac{(\partial_b\delta^{(0)})^4}{p_2^+ p_1^-} = O(\gamma^{-5}). \label{OnShell Condition}
\end{align}
All the $f$ and $g$ functions have canceled, leaving only terms at fixed PM order. This is a non-trivial check on the resummed expressions; moreover, we may use a Eq. (\ref{OnShell Condition}) to fix $\Delta p^+_{1,SL}$, which then gives the next-to-leading minus impulse at 2PM and leading minus impulse at 4PM. Comparing to the explicit results from \cite{Dlapa:2022lmu}, we find complete agreement. 

As a secondary cross-check on the amplitudes used in the computation in Eqs. (\ref{4-Point}) and (\ref{5-point}), we may compute the radiated energy spectrum of a collinear graviton in the case of massless scattering. This may be done by, e.g., using Eq. (\ref{Wave Impulse}) for the impulse and freezing the $y$-integration. We find
\begin{widetext}
\begin{equation}
   \frac{d E_{\text{Rad}}}{dy} =2\frac{\sqrt{s}\, G_N}{\pi^2} \theta(-y)\theta(1 + y)\frac{(1 + y)}{y^2}\int \frac{d^2b^\prime}{{b^\prime}^4}\left[2-\frac{(b^2)^{-iG_N s}[(b+y b^\prime)^2]^{i G_N s( 1+y)}}{[(b + (1 +y)b^\prime)^2]^{i G_N s y}}-\frac{[(b^2)^{i G_N s}(b+y b^\prime)^2]^{-i G_N s( 1+y)}}{[(b + (1 +y)b^\prime)^2]^{-i G_N s y}}\right].\label{Massless Quantum Impulse}
\end{equation}
\end{widetext}
It is not obvious that we are able to take the classical limit of this expression, since $y\sim 1$. However, if we consider the limit of small $y$, with $y\sim 1/(G_N s)$ as in \cite{Gruzinov:2014moa}, we find
\begin{equation}
  \frac{d E_{\text{Rad}}}{dy} \simeq 
  \frac{8 }{\pi^2}\frac{G_N\sqrt{s}}{y^2}\int\frac{d^{d^\prime}b^\prime}{(b^\prime)^4}\sin^2\left(\frac{y\Phi(b^\prime)}{2}\right).
\end{equation}
This is exactly the result found in \cite{Gruzinov:2014moa, Ciafaloni:2015xsr} for the classically radiated energy spectrum of massless particles.  
It is rather interesting to note that the total energy radiated in the massless case is divergent in the UV \cite{Gruzinov:2014moa}, while in the massive case it is finite. It seems strange that the presence of a mass term leads to different UV behavior; typically, one expects the UV behavior of a theory to be independent of IR parameters such as the mass.

\paragraph*{Discussion}

Using Eq. (\ref{Radiated Momentum}), we find that the total radiated momentum is given as
\begin{equation}
R^\mu =  -\frac{\bar{n}^\mu}{2p_2^+}f^+(m_1) -\frac{n^\mu}{2p_1^-}f^+(m_2) + O(\gamma^{-4}).\label{RadiatedMomentum}
\end{equation}
As the high-energy coupling becomes large, the radiated energy behaves asymptotically as
\begin{equation}
   R^0 = E_{\text{Rad}}\xrightarrow{\alpha\rightarrow \infty} \frac{2}{\pi}\frac{m_1 m_2}{\sqrt{s}} {\alpha}^2 \log \alpha .
\end{equation}

An important caveat is that this calculation breaks down for fixed-angle scattering. Since $\alpha \sim \theta \gamma$, where $\theta \sim G_N \sqrt{s}/b$ is the scattering angle, one encounters terms which scale as $\gamma^{-2}\alpha^2\sim \theta^2$, indicating a breakdown of the power-counting. This is made manifest in the radiated energy, which then suffers from a logarithmic divergence at high energies:
\begin{equation}
E_{\text{Rad}}\xrightarrow{s/m^2\rightarrow \infty} \frac{2}{\pi}\sqrt{s}\,\theta^2 \log\left(\theta\frac{\sqrt{s}}{m}\right).
\end{equation}
This is in contrast to the expectations from the massless case or the ZFL of massive scattering, where the radiated energy behaves as $E_{\text{Rad}}\sim \sqrt{s}\,\theta^2\log\theta$ \cite{Gruzinov:2014moa, Ciafaloni:2015xsr, Ciafaloni:2018uwe, DiVecchia:2022nna, Alessio:2024onn} (See also \cite{Damour:2019lcq}).  Therefore, we are still restricted to a vanishingly small scattering angle at high energies $\theta\sim \sigma^{-\varepsilon}$ for $\varepsilon >0$, or equivalently $b\sim \sigma^{1/2 + \varepsilon}$, though this is a much softer scaling than the D'Eath bound of $\theta \sim \sigma^{-1/2}$ ($b\sim \sigma$).

To clarify this discussion, we consider relative scattering angle of particle 1.  To $O(\gamma^{-5})$, this is given as
\begin{align}
    \theta_1 =&\theta^{(1\text{PM})}-\frac{(\theta^{(1\text{PM})})^3}{12} + \frac{m_1^2+ m_2^2}{s}\theta^{(1\text{PM})} + \frac{2\alpha m_1}{s^{3/2}}f^-\nonumber\\
    &  - \frac{15\pi m_1^3}{4 s^{3/2}}\alpha^3\left(1 + \frac{m_2}{m_1}\right) + \frac{2}{\sqrt{s}}\hat{b}\cdotp f_\perp  + O(\gamma^{-5}),\label{scattering angle}
\end{align}
where $\theta^{(1\text{PM})} = 4 G \sqrt{s}/b\sim \gamma^{-1}$ is the 1PM scattering angle at high-energies.  This dominates the scattering angle, as all other terms scale as $\sim \gamma^{-3}$.  It also makes manifest, however, that at sufficiently large $\alpha$ the power-counting breaks down.  Keeping $\theta^{(1\text{PM})}\sim \theta$ fixed one then finds that the scattering angle grows at large $s/m^2$ as
\begin{equation}
    \theta_1 \sim \theta + \theta^3 \log\left(\theta\frac{\sqrt{s}}{m}\right) + ....
\end{equation}
The $...$ contains terms from the $O(\gamma^{-5})$ in Eq. (\ref{scattering angle}), which may contain higher powers of logs.  The important point is that we can explicitly see terms which are subleading in $m^2/s$ at fixed $\alpha$ now enter at leading power at fixed $\theta$. The scattering angle then develops the log divergence, indicating a breakdown of the computation in this regime.


\paragraph*{Conclusions}

We have presented an all orders in $G_N$ resummed formula for the classical gravitational impulse, allowing for one to probe the case of strongly coupled scattering with $G_N s/(m_1 b)\sim 1$. Using this formula, we gave predictions for the leading high-energy contribution to the PM-expanded impulse through $O(G_N^{11})$. We found agreement with all PM results through 4PM \cite{Dlapa:2022lmu, Driesse:2024feo} along with a resummed result for the radiated energy spectrum in the massless limit \cite{Gruzinov:2014moa}. 

There are several interesting avenues for further study. Most notably, we are restricted to $G_N s/(m b) \sim 1$, and so gravitational scattering at arbitrarily high energies is still out of reach. Due to the power-counting breakdown, it seems different techniques will be needed to access this region of phase space. Pushing this idea to higher orders in $1/\sigma$ seems daunting, given the need to sum over all possible Glauber exchanges. One possible solution is to develop an EFT for this regime using some idea close to that of \cite{Cheung:2023lnj, Cheung:2024byb}. In principle, this could allow one to work with the shockwave propagator \cite{Raj:2023irr} directly instead of having to sum over Glauber exchanges by hand. Additionally, this analysis ignores both Love tidal effects and spin \cite{Goldberger:2005cd,Porto:2016pyg, Vines:2017hyw}, which may play a role at high energies \cite{ Alessio:2026bdi}; it would be interesting to see if these lead to non-trivial effects. Similarly, we have not considered static modes \cite{DiVecchia:2022owy,Manohar:2022dea, Bini:2024rsy,Elkhidir:2024izo, Heissenberg:2024umh}, which do not play a role in the impulse, but are important for related observables such as the radiated energy spectrum \cite{DiVecchia:2022nna}. Lastly, it would be interesting to apply similar techniques to compute other observables, such as the scattering waveform \cite{Cristofoli:2021vyo} or energy correlators \cite{Herrmann:2024yai}.

\begin{acknowledgments}
We thank Zvi Bern, Vittorio Del Duca, Enrico Herrmann, Ira Rothstein, and Michael Ruf for useful comments and discussions during this work, and to Thibault Damour and Ira Rothstein for helpful comments on the draft. We also thank Ryan Plestid for work on a related project. Feynman diagrams were made using \cite{Ellis:2016jkw}.  M.S. is supported in part by the U.S. Department of Energy (DOE) under award number DE SC0009937, and by the European Research Council (ERC) Horizon Synergy Grant “Making Sense of the Unexpected in the Gravitational Wave Sky” grant agreement no. GWSky–101167314. We are also grateful to the Mani L. Bhaumik Institute for Theoretical Physics for support.
\end{acknowledgments}

\appendix

\section{4- and 5-Point Amplitudes with Collinear Gravitons}\label{Appendix_Amps}

\begin{figure}
\begin{subfigure}[b]{.15\textwidth}
  \scalebox{.7}{
\begin{tikzpicture}
\begin{feynman}
	\vertex [label = below: \(1\)] (i1) at (-1.75,1);
	\vertex [label = below: \(\)] (f1) at (1.75, 1);
	\vertex [label = above: \(2\)] (i2) at (-1.75,-1);
	\vertex [label = above: \(\)] (f2) at (1.75, -1);
	\node[dot] (cn1) at (-1.25,1);
	\node[dot] (cn2) at (1.25,1);
  \vertex (c3) at (0, .55);
	\vertex (c4) at (0, -1);
  \vertex (g11) at (-.9,1);
  \vertex (g21) at (-.9,-1); 
  \vertex (g12) at (-.6,.55);
  \vertex (g22) at (-.6,-1); 
  \vertex (g13) at (-.3,.48);
  \vertex (g23) at (-.3,-1); 
  \vertex (g14) at (.3,1);
  \vertex (g24) at (.3,-1); 
  \vertex (g15) at (.6,.55);
  \vertex (g25) at (.6,-1); 
  \vertex (g16) at (.9,1);
  \vertex (g26) at (.9,-1); 
    \vertex[label = : \(\color{red}{...}\)] at (0,-.2);
\diagram*{
	(i1)--[scalar, line width = 0.3mm](cn1)--[scalar, line width = 0.3mm](cn2)--[scalar, line width = 0.3mm](f1),
	(f2)--[scalar, line width = 0.3mm](c4)--[scalar, line width = 0.3mm](i2),
  (cn1)--[line width = 0.3mm, quarter right, looseness = 1](cn2),
  (cn1)--[gluon, line width = 0.2mm, quarter right, looseness = 1](cn2),
  (g11)--[scalar, line width = 0.3mm, red](g21),
  (g12)--[scalar, line width = 0.3mm, red](g22),
  (g13)--[scalar, line width = 0.3mm, red](g23),
  (g14)--[scalar, line width = 0.3mm, red](g24),
  (g15)--[scalar, line width = 0.3mm, red](g25),
  (g16)--[scalar, line width = 0.3mm, red](g26),
};
\end{feynman}
  \filldraw[red] (-.9,1) ellipse (0.6mm and 1.2mm);
  \filldraw[red] (-.6,.55) ellipse (0.6mm and 1.2mm);
  \filldraw[red] (-.3,.48) ellipse (0.6mm and 1.2mm);
  \filldraw[red] (.3,1) ellipse (0.6mm and 1.2mm);
  \filldraw[red] (.6,.55) ellipse (0.6mm and 1.2mm);
  \filldraw[red] (.9,1) ellipse (0.6mm and 1.2mm);
  \filldraw[red] (-.9,-1) ellipse (0.6mm and 1.2mm);
  \filldraw[red] (-.6,-1) ellipse (0.6mm and 1.2mm);
  \filldraw[red] (-.3,-1) ellipse (0.6mm and 1.2mm);
  \filldraw[red] (.3,-1) ellipse (0.6mm and 1.2mm);
  \filldraw[red] (.6,-1) ellipse (0.6mm and 1.2mm);
  \filldraw[red] (.9,-1) ellipse (0.6mm and 1.2mm);
\end{tikzpicture}
}
\end{subfigure}
\begin{subfigure}[b]{.15\textwidth}
\scalebox{0.7}{
\begin{tikzpicture}
\begin{feynman}
	\vertex (f1) at (-.5,1);
	\vertex (f2) at (-.5,-1);
	\node[dot] (k1) at (-.75,1);
	\vertex[label = above : \(\ell\)] (k2) at (1.5, 1.5);
	\vertex (p3) at (1.5, .5);
	\vertex[label = below : \(1\)] (p2) at (-1.5, 1);
  \vertex[label = above : \(2\)] (p1) at (-1.5, -1);
	\vertex(p4) at (1.5, -1);
  \vertex(g11) at (-.2,1.1);
  \vertex(g21) at (-.2,-1);
  \vertex(g12) at (0,.85);
  \vertex(g22) at (0,-1);
  \vertex(g13) at (.2,1);
  \vertex(g23) at (.2,-1);
  \vertex(g14) at (.7,1.33);
  \vertex(g24) at (.7,-1);
  \vertex(g15) at (.9,1.38);
  \vertex(g25) at (.9,-1);
  \vertex(g16) at (1.1,.57);
  \vertex(g26) at (1.1,-1);
  \vertex[label = : \(\color{red}{...}\)] at (.45,-.2);
\diagram*{
	(p2)--[scalar, line width = 0.3mm](f1)--[scalar, line width = 0.3mm](k1)--[scalar, line width = 0.3mm](p3),
	(p4)--[scalar, line width = 0.3mm](p1),
	(k1)--[gluon, line width = 0.2mm](k2),
	(k1)--[ line width = 0.2mm](k2),
  (g11)--[scalar, line width = 0.3mm, red](g21),
  (g12)--[scalar, line width = 0.3mm, red](g22),
  (g13)--[scalar, line width = 0.3mm, red](g23),
  (g14)--[scalar, line width = 0.3mm, red](g24),
  (g15)--[scalar, line width = 0.3mm, red](g25),
  (g16)--[scalar, line width = 0.3mm, red](g26),
};
\end{feynman}
  \filldraw[red] (-.2,1.1) ellipse (0.6mm and 1.2mm);
  \filldraw[red] (0,.85) ellipse (0.6mm and 1.2mm);
  \filldraw[red] (.2,1.2) ellipse (0.6mm and 1.2mm);
  \filldraw[red] (.7,1.33) ellipse (0.6mm and 1.2mm);
  \filldraw[red] (.9,1.38) ellipse (0.6mm and 1.2mm);
  \filldraw[red] (1.1,.57) ellipse (0.6mm and 1.2mm);
  \filldraw[red] (-.2,-1) ellipse (0.6mm and 1.2mm);
  \filldraw[red] (0,-1) ellipse (0.6mm and 1.2mm);
  \filldraw[red] (.2,-1) ellipse (0.6mm and 1.2mm);
  \filldraw[red] (.7,-1) ellipse (0.6mm and 1.2mm);
  \filldraw[red] (.9,-1) ellipse (0.6mm and 1.2mm);
  \filldraw[red] (1.1,-1) ellipse (0.6mm and 1.2mm);
\end{tikzpicture}
}
\end{subfigure}
\begin{subfigure}[b]{.15\textwidth}
\scalebox{0.7}{
\begin{tikzpicture}
\begin{feynman}
	\vertex (f1) at (-.5,1);
	\vertex(f2) at (-.5,-1);
	\node[dot] (k1) at (.25,1);
	\vertex[label = above : \(\ell\)] (k2) at (1.5, 1.5);
	\vertex (p3) at (1.5, .5);
	\vertex[label = below : \(1\)] (p2) at (-1.5, 1);
  \vertex[label = above : \(2\)] (p1) at (-1.5, -1);
	\vertex(p4) at (1.5, -1);
  \vertex(g11) at (-.1,1);
  \vertex(g12) at (-.3,1);
  \vertex(g13) at (-.5,1);
  \vertex(g14) at (-1,1);
  \vertex(g15) at (-1.2,1);
  \vertex(g21) at (-.1,-1);
  \vertex(g22) at (-.3,-1);
  \vertex(g23) at (-.5,-1);
  \vertex(g24) at (-1,-1);
  \vertex(g25) at (-1.2,-1);
  \vertex[label = : \(\color{red}{...}\)] at (-.75,-.1);
\diagram*{
	(p2)--[scalar, line width = 0.3mm](f1)--[scalar, line width = 0.3mm](k1)--[scalar, line width = 0.3mm](p3),
	(p4)--[scalar, line width = 0.3mm](p1),
	(k1)--[gluon, line width = 0.2mm](k2),
	(k1)--[ line width = 0.2mm](k2),
  (g11)--[scalar, red, line width = 0.3mm](g21),
  (g12)--[scalar, red, line width = 0.3mm](g22),
  (g13)--[scalar, red, line width = 0.3mm](g23),
  (g14)--[scalar, red, line width = 0.3mm](g24),
  (g15)--[scalar, red, line width = 0.3mm](g25),

};
\end{feynman}
  \filldraw[red] (-.5,1) ellipse (0.6mm and 1.2mm);
  \filldraw[red] (-.3,1) ellipse (0.6mm and 1.2mm);
  \filldraw[red] (-.1,1) ellipse (0.6mm and 1.2mm);
  \filldraw[red] (-1,1) ellipse (0.6mm and 1.2mm);
  \filldraw[red] (-1.2,1) ellipse (0.6mm and 1.2mm);
  \filldraw[red] (-.5,-1) ellipse (0.6mm and 1.2mm);
  \filldraw[red] (-.3,-1) ellipse (0.6mm and 1.2mm);
  \filldraw[red] (-.1,-1) ellipse (0.6mm and 1.2mm);
  \filldraw[red] (-1,-1) ellipse (0.6mm and 1.2mm);
  \filldraw[red] (-1.2,-1) ellipse (0.6mm and 1.2mm);
\end{tikzpicture}
}
\end{subfigure}
\caption{Diagrams for the 4 and 5 point amplitudes in massive SCET. Here, the red dashed lines indicate the exchange of a Glauber graviton.  There are an equivalent set of diagrams with $1\leftrightarrow 2$ and $n\leftrightarrow \bar{n}$.}
  \label{fig:AmpDiagrams}
\end{figure}
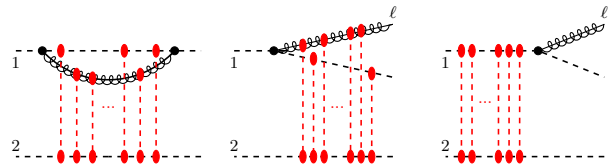

Here we present the results for the contributions to the 4- and 5-point amplitudes relevant for the computation of the classical impulse. We perform this computation within the EFT framework of \cite{Rothstein:2024nlq}, generalized slightly to allow for massive collinear projectiles. The details of this computation will be given in a forthcoming publication \cite{Forthcoming}. Here we give only the terms in the amplitude involving collinear gravitons. While the amplitudes involving soft gravitons do appear at this order, their contributions cancel in the sum of the real and virtual terms.

For the 4-point amplitude, the relevant class of diagrams is the ``rainbow diagrams", which contain one collinear graviton emission off of a collinear scalar, with an arbitrary number of Glauber exchanges. We take $\ell$ to be the momentum of the collinear graviton, and we define $y = \ell^- /p_1^-$. The sum of these diagrams is then given as
\begin{align}
  \frac{\tilde{\mathcal{M}(b)}}{2s} =& \frac{1}{\pi}\int_{-1}^0 \frac{(\nu/p_1^-)^\eta\,d y}{(1 + y)(-y)^{1 + \eta}} \int d^{d^\prime} b^\prime N (b^\prime)\label{4-Point}\\
  &\times\left(e^{i \delta_y(b,b^\prime) + i\delta^{(0)}(b) } -1\right) + (1\leftrightarrow 2),\nonumber
\end{align}
where $b^\prime$ is the Fourier dual of $\ell_\perp$. $\delta_y$ is the IR-finite function
\begin{equation}
  \delta_y(b,b^\prime) = (1 + y)\delta^{(0)}(b - y b^\prime) - y \,\delta^{(0)}(b -(1 + y)b^\prime) - \delta^{(0)}(b),
\end{equation}
and $\eta$ is a rapidity regulator needed to make the integral over $y$ well-defined \cite{Chiu:2011qc, Chiu:2012ir}. $N(b^\prime)$ is the Fourier transform of the $\ell_\perp$ integral, and it is given as
\begin{align}
  N(b^\prime) &= \tilde{V}^{\mu\nu}(b^\prime)P_{\mu\nu,\rho\sigma}\tilde{V}^{\rho\sigma}(b^\prime),\nonumber\\
  \tilde{V}^{\mu\nu}(b^\prime) &= \frac{\kappa\sqrt{2}(1 + y)}{y}\int [d^{d^\prime}k_\perp] \frac{k^\mu_\perp k^\nu_\perp\, e^{i b^\prime\cdotp k}}{\vec{k}^{\,2} + y^2 m_1^2}.\label{4-Point Vertex Function}
\end{align}
Here, we have neglected contributions from soft loops, which do not contribute to the impulse at the working order. We have also dropped terms which lead to integrals that vanish in either the heavy particle ($m_j\rightarrow \infty$) or the massless limits. These are tadpole contributions which involve integrals of the form
\begin{equation}
  I_{tadpole} \sim \int_{-1}^0 \frac{dy}{(-y)^{2 + \eta}}\int \frac{[d^{d^\prime}\ell_\perp]}{\ell_\perp^2 - y^2 m_j^2}f(\ell_\perp, y).
\end{equation}
In the massless limit, the $\ell_\perp$ integral is scaleless, and in the heavy particle limit the integral also becomes scaleless as the integration domain of $y$ is extended from $[-1,0]$ to $(-\infty, 0]$. Such contributions arise from either wave-function renormalization contributions or graphs involving Wilson line emission off of the scalar-scalar Glauber vertex. 

There are some crucial simplifications we may make in the computation of the 5-point amplitude for the emission of a collinear graviton. In particular, using the graviton equations of motion $\ell^\mu \epsilon_{\mu\nu} = 0$, we may remove all occurrences of $\epsilon_{+\mu}$ using
\begin{equation}
  \epsilon_{+\mu} = \frac{\ell_\perp^2}{(\ell^-)^2}\epsilon_{-\mu} - 2\frac{\ell_\perp^\nu \epsilon^\perp_{\nu\mu}}{\ell^-}.
\end{equation}
The full 5-point amplitude then only depends on $\epsilon_{--}$, $\epsilon_{\mu-}^\perp$, and $\epsilon_{\mu\nu}^{\perp\perp}$. Next, we use the fact that gauge invariance of the amplitude allows us to then write the sum over polarizations as
\begin{equation}
  \sum_{\text{pols}}\epsilon_{\mu\nu}(\ell)\epsilon_{\rho\sigma}(-\ell) \rightarrow P_{\mu\nu,\rho\sigma}.
\end{equation}
Since we have removed all $+$ polarizations, and term proportional to $\epsilon_{-\mu}$ then drops out in the polarization sum, and the only contribution comes from terms with (the physical) transverse polarizations. A second important simplification is the Glauber collapse rule\cite{Rothstein:2016bsq}, which states that a diagram must allow for all Glauber exchanges to slide on top of each other without impediment from vertices. This restricts the class of diagrams to then be those in which the Glauber exchanges happen either entirely before or after the scalar-graviton vertex. Diagrams involving Wilson line emission may be discarded since they are proportional to a $-$ polarized graviton $\epsilon_{-\mu}$.
The transverse piece of the 5-point amplitude may then be written as
\begin{align}
  \tilde{\mathcal{M}}_5^{\perp\perp}(\ell_\perp, b) =& \int d^{d^\prime}b^\prime \tilde{V}(b^\prime) e^{i b^\prime\cdotp (\ell_\perp - y p_{1\perp})}e^{i\delta^{(0)}(b +y b^\prime)}\nonumber\\
  &\times \left(e^{i \delta_y(b + y b^\prime,b^\prime) } -1\right).\label{5-point}
\end{align}
Here, we have defined $s = p_1^- p_2^+$ as the center-of-mass energy of the incoming states, $-\ell$ is the momentum of the radiated graviton, and $y = \ell^-/p_1^-$. 
The vertex function $\tilde{V}$ is similar to the one in Eq. (\ref{4-Point Vertex Function}), and is given as
\begin{equation}
  V(b^\prime) = 2\sqrt{2}\frac{\kappa\,s(1 + y)}{\,y |y|^{\eta/2}}\left(\frac{\nu}{p_1^-}\right)^{\eta/2}\int [d^{d^\prime}k_\perp] \frac{k^\mu_\perp k^\nu_\perp\,\epsilon_{\mu\nu}^{\perp\perp}(k)}{\vec{k}_\perp^{\,2} + y^2 m_1^2} e^{i b^\prime\cdotp k}.
\end{equation}
We can see that the IR divergences explicitly factor out into an overall phase, as dictated by the Weinberg soft theorems at high energies \cite{Weinberg:1965nx}. When we compute the impulse with these amplitudes, the results are finite as $y\rightarrow 0$, so we may set the rapidity regulator $\eta$ to zero, with the exception of $g_\perp$ in Eq. (\ref{gperp}).

\newpage

\onecolumngrid

\section{The Impulse to $O(G_N^{11})$}\label{B}

In this appendix, we list the perturbative expansions for the impulse through $O(G_N^{11})$.  In particular, we give the transverse impulse at $\gamma^{-1}$ and $\gamma^{-3}$,  and the plus impulse at $\gamma^{-2}$ and $\gamma^{-4}$.  The minus impulse at $\gamma^{-4}$ is not computed in this work, following the discussion above Eq. (\ref{N3LO Impulse}).  As above, we use $l_4 \equiv \log(4)$, and $\alpha_1 = G_N s/m_1 b$.

\begin{align}
    \hat{b}\cdotp\Delta p_{1\perp,(1)} =& -2m_1\alpha_1 \\
    \frac{s}{m_1^3}\hat{b}\cdotp\Delta p_{1\perp,(3)} =&  -\frac{15\pi}{8}\alpha_1^2\left(1 + \frac{m_2}{m_1}\right) + \frac{35\pi\alpha_1^4}{64}\left(-3l_4^2 + 4l_4 +7\right) + \frac{315\pi\alpha_1^6}{32}\bigg(\frac14 l_4^4-\frac{26}{15}l_4^3 + \frac{27}{10}l_4^2\nonumber\\
    & + \left(3\zeta_3-\frac{11}{5}\right)l_4-\frac{26}{5}\zeta_3 + \frac{103}{60}\bigg) + \frac{495\pi\alpha_1^8}{2^{11}}\bigg(-\frac74 l_4^6 + \frac{113}{5}l_4^5 - \frac{853}{12}l_4^4 + \left(-105\zeta_3 + \frac{103}{3}\right)l_4^3 \nonumber\\
    & + \left(678\zeta_3 + \frac{171}{4}\right)l_4^2 + \left(-\frac{945}{2}\zeta_5 -853\zeta_3-67\right)l_4 -\frac{315}{2}\zeta_3^2+ 1017\zeta_5 + 103 \zeta_3 + \frac{3851}{60}\bigg)\\
    & +\frac{45045\pi\alpha_1^{10}}{2^{15}}\bigg(\frac14 l_4^8 -\frac{1532}{315}l_4^7+\frac{18053}{675}l_4^6  +  \left[42\zeta_3-\frac{29342}{675}\right]l_4^5 + \left[-\frac{1532}{3}\zeta_3 + \frac{5021}{270}\right]l_4^4\nonumber\\
    &+  \left[630\zeta_5+ \frac{72212}{45}\zeta_3 - \frac{2092}{135}\right]l_4^3
     + \left[630\zeta_3^2-4596\zeta_5- \frac{58684}{45}\zeta_3 + \frac{1657}{45}\right]l_4^2 +\bigg[2835\zeta_7 - 3064 \zeta_3^2 \nonumber\\
     &+ \frac{36106}{5}\zeta_5 + \frac{10042}{45}\zeta_3 - \frac{3182}{45}\bigg]l_4 +1890\zeta_5\zeta_3 -6894\zeta_7 + \frac{36106}{15}\zeta_3^2 - \frac{29342}{15}\zeta_5 -\frac{2092}{45}\zeta_3 + \frac{440719}{6300}
    \bigg) \nonumber
\end{align}

\begin{align}
    \frac{p_1^-}{m_1^2}\Delta p_{1,(2)}^{+} =& 4\alpha_1^2
    \\
    \frac{s p_1^-}{m_1^4}\Delta p_{1,(4)}^{+} =& 4\alpha_1^2+\pi\alpha_1^3\left(\frac{15}{2}\left(1 + \frac{m_1}{m_2}\right) + \frac{35}{32}(1 + l_4)\right)  + 16 \alpha_1^4-\frac{7\pi\alpha_1^5}{128}\left(9l_4^3 - 147l_4^2 + 107l_4 + 27\zeta_3 + 182\right)\nonumber\\
    &+ \frac{1485\pi\alpha_1^7}{2048}\left(\frac{l_4^5}{2} - \frac{499}{66}l_4^4 + \frac{1481}{55}l_4^3 + \left[15\zeta_3-\frac{4154}{165}\right]l_4^2 + \left[-\frac{998}{11}\zeta_3 + \frac{1931}{110}\right]l_4 + \frac{45}{2}\zeta_5 + \frac{4443}{55}\zeta_3 - \frac{13901}{990}\right)\nonumber\\
    & - \frac{5005\pi\alpha_1^9}{16384}\bigg(l_4^7 - \frac{3901}{195}l_4^6 + \frac{795281}{6825}l_4^5 +\left[105\zeta_3 - \frac{290026}{1365}\right]l_4^4+\left[-\frac{15604}{13}\zeta_3 + \frac{88867}{1365}\right]l_4^3\nonumber\\ & +\left[945\zeta_5 + \frac{1590562}{455}\zeta_3 + \frac{35057}{455}\right]l_4^2+ \left[630 \zeta_3^2 -\frac{70218}{13}\zeta_5 - \frac{1160104}{455}\zeta_3 - \frac{11237}{91}\right]l_4 + \frac{2835}{2}\zeta_7\nonumber\\
    & -\frac{23406}{13}\zeta_3^2+ \frac{2385843}{455}\zeta_5 + \frac{88867}{455}\zeta_3 + \frac{54024}{455}\bigg) + \frac{143325\pi\alpha^{11}}{65536}\bigg(-\frac{l_4^9}{8} + \frac{741}{280}l_4^8 -\frac{38327}{2450}l_4^7\nonumber\\
    & + \left[-\frac{63}{2}\zeta_3 + \frac{4561}{175}\right]l_4^6 + \left[\frac{2223}{5}\zeta_3-\frac{7017}{700}\right]l_4^5 + \left[-\frac{2835}{4}\zeta_5 - \frac{114981}{70}\zeta_3 - \frac{537}{140}\right]l_4^4 + \bigg[-945\zeta_3^2\nonumber\\
    &+ 6669\zeta_5 + \frac{54732}{35}\zeta_3 + \frac{129}{14}\bigg]l_4^3 + \bigg[-\frac{25515}{4}\zeta_7  + 6669\zeta_3^2- \frac{1034829}{70}\zeta_5-\frac{21051}{70}\zeta_3 - \frac{888}{35}\bigg]l_4^2\nonumber\\
    & + \bigg[-8505\zeta_5\zeta_3 + \frac{60021}{2}\zeta_7 -\frac{344943}{35}\zeta_3^2 + \frac{246294}{35}\zeta_5 - \frac{1611}{35}\zeta_3 + \frac{13  901}{280}\bigg]l_4-\frac{80325}{8}\zeta_9 -945 \zeta_3^3\nonumber\\
    &+ 20007\zeta_5\zeta_3 - \frac{3104487}{140}\zeta_7 + \frac{82098}{35}\zeta_3^2 - \frac{65153}{140}\zeta_5 + \frac{387}{14}\zeta_3 - \frac{207181}{4200}\bigg).
\end{align}

\twocolumngrid
\bibliography{references}

\end{document}